\documentclass[
reprint,
twocolumn,
floatfix,
longbibliography,
superscriptaddress,
amssymb,
amsmath
]{revtex4-2}

\usepackage[dvips]{graphicx}
\usepackage{epsf}
\usepackage{dcolumn}
\usepackage{bm}
\usepackage{bm, color}
\usepackage{bbm}
\usepackage{lineno}
\usepackage{braket}
\usepackage{amsbsy}
\usepackage{xr}
\usepackage{easyReview}

\usepackage[
colorlinks=true,
linkcolor=magenta,
urlcolor=magenta,
citecolor=magenta,
filecolor=blue,
pagecolor=blue,
linktocpage=true,
bookmarks=true,
bookmarksnumbered=true
]{hyperref}

\begin{document}

	\title{Electric field tunable non-linear Hall terahertz detector in Dual quantum spin Hall insulator $\text{TaIrTe}_4$}
	
	\author{Junwen Lai}
	\affiliation{School of Materials Science and Engineering, University of Science and Technology of China, Shenyang 110016, China.}
	\affiliation{Shenyang National Laboratory for Materials Science, Institute of Metal Research,Chinese Academy of Sciences, Shenyang 110016, China.}

    \author{Jie Zhan}
    \affiliation{School of Materials Science and Engineering, University of Science and Technology of China, Shenyang 110016, China.}
    \affiliation{Shenyang National Laboratory for Materials Science, Institute of Metal Research,Chinese Academy of Sciences, Shenyang 110016, China.}

	\author{Peitao Liu}
	\affiliation{School of Materials Science and Engineering, University of Science and Technology of China, Shenyang 110016, China.}
	\affiliation{Shenyang National Laboratory for Materials Science, Institute of Metal Research,Chinese Academy of Sciences, Shenyang 110016, China.}
	
	\author{Xing-Qiu Chen}
	\email{xingqiu.chen@imr.ac.cn}
	\affiliation{School of Materials Science and Engineering, University of Science and Technology of China, Shenyang 110016, China.}
	\affiliation{Shenyang National Laboratory for Materials Science, Institute of Metal Research,Chinese Academy of Sciences, Shenyang 110016, China.}
	
	\author{Yan Sun}
	\email{sunyan@imr.ac.cn}
	\affiliation{School of Materials Science and Engineering, University of Science and Technology of China, Shenyang 110016, China.}
	\affiliation{Shenyang National Laboratory for Materials Science, Institute of Metal Research,Chinese Academy of Sciences, Shenyang 110016, China.}

	\begin{abstract}
		Nonlinear Hall effect (NHE) can be generated via Berry 
		curvature dipole (BCD) on nonequilibrium Fermi surface 
		in a non-magnetic system without inversion symmetry.
		To achieve a large BCD, strong local Berry curvatures 
		and their variation with respect to momentum are necessary 
		and hence topological materials with strong inter-band 
		coupling emerge as promising candidates. In this study, 
		we propose a switchable and robust BCD in the newly
		discovered dual quantum spin Hall insulator (QSHI) $\text{TaIrTe}_4$ 
		by applying out-of-plane electric fields. Switchable BCD 
		could be found along with topological phase transitions or 
		insulator-metal transition in the primitive cell and CDW phases of 
		$\text{TaIrTe}_4$ monolayer. This work presents an 
		instructive strategy for achieving a switchable and robust 
		BCD within dual QSHIs, which should be helpful for designing 
		the NHE-based THz radiations detector.
	\end{abstract}
	
	\maketitle
	\section{Introduction}
\begin{figure}[htbp]
		\begin{center}
			\includegraphics[width=0.40\textwidth]{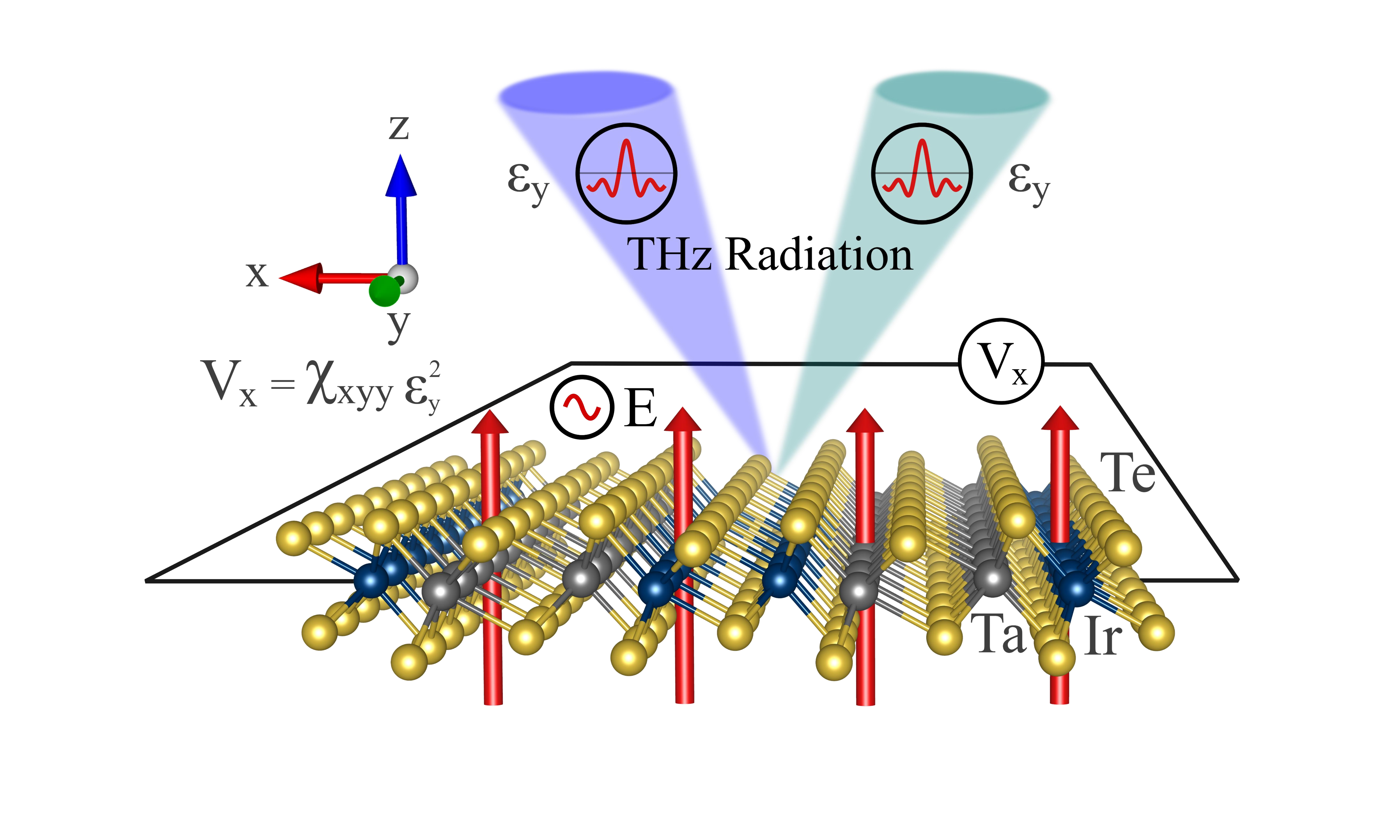}
		\end{center}
		\caption{
			Schematic view of terahertz (THz) photodetection based on 
			the Berry curvature dipole (BCD) of the quantum spin Hall 
			insulator (QSHI) $\text{TaIrTe}_4$.	$\varepsilon_y$ is 
			the amplitude of the THz polarized light along $y$ direction,
			$j_x$ is the non-linear DC along $x$ direction,
			$E$ is the electric field perpendicular to the monolayer 
			$\text{TaIrTe}_4$. The strength of this mechanism could be 
			described by the nonlinear conductivity $\chi_{xyy}$.
		}
		\label{fig1}
	\end{figure}

    In nonlinear second-order quantum responses, the Berry 
	curvature dipole (BCD) has recently attracted much 
	attention as it could widely occur in non-magnetic 
	systems with breaking inversion symmetry at both 
	zero and twice driving frequency\cite{sodemann2015,du2021}.
	Unlike the well-known anomalous Hall effect (AHE) which 
	serves as a summation of Berry curvature, BCD manifests 
	as a spatially varying distribution of Berry curvature 
	in momentum space and does not require time reversal 
	symmetry broken. In the last decades, BCD has provided 
	inspirations in many physical properties and 
	applications including spintronics, energy harvesting and 
	microwave detection 
	\cite{xiao2010,you2018,zhang2018a,zhang2018,xu2018,wang2021,
	pantaleon2021,zhang2021,sinha2022,fu2023,ye2023,duan2023,
	lempicka-mirek}.
	Very recently, the utilization of BCD as a rectifier 
	of terahertz (THz) wave provides a new strategy for 
	the THz radiation detection based on intrinsic 
	electronic band structures in materials\cite{zhang2021}, see Fig.~\ref{fig1}. 
	The THz technology is making groundbreaking advancement and has already 
	been proposed as a next-generation 
	technology in many fields including spectroscopy, sensing, 
	information and communication\cite{tonouchi2007,stantchev2020}. 
	Despite these wide potentials, its practical application 
	as THz detection still faces challenges in obtaining a 
	sensitive, switchable, and significant signal, as well as 
	the compatibility to integral into currently using electronic 
	devices. Therefore, new design of the devices and materials
	candidates are desired.

	Materials with high BCD have been studied in various systems 
	including topological insulators (TIs), Weyl semimetals 
	(WSMs), transition metal dichalcogenides, anti-ferromagnetic 
	metals, etc., both theoretically and experimentally 
	\cite{you2018,zhang2018a,zhang2018,xu2018,wang2021,pantaleon2021,
	zhang2021,sinha2022,fu2023,ye2023,duan2023,lempicka-mirek}. 
	Topological materials are naturally with great potential to 
	have large BCD due to their large local Berry curvature originating 
	from the band inversion. The electric field serves as an effective 
	method to tune the electronic structures and induce phase transition 
	in two-dimensional (2D) topological materials\cite{maciejko2011,wu2018,xu2018,ye2023,lempicka-mirek}.
	Therefore, tunable NHE in 2D topological materials is expected in 
	the condition with gating fields. The experimental studies about 
	BCD in 2D TIs are rare due to the lack of promising material candidates.
	Most studies focus on transition metal dichalcogenides $\text{WTe}_2\text{/MoTe}_2$
	and the moiré superlattice\cite{you2018,zhang2018, xu2018,ye2023}.

	Very recently, a new topological phase of dual quantum spin Hall 
	insulator (QSHI) was discovered in $\text{TaIrTe}_4$\cite{tang2024} 
	monolayer. In addition to the nontrivial band gap at the charge 
	neutrality points\cite{liu2017a,guo2020}, a new $Z_2$ gap can be 
	generated when the chemical potential shifts to the van Hove 
	singularity (VHS) point, which reveals strong inter-band coupling 
	in a large energy window. In this work, the electronic structure 
	and topological phase transition of this new phase are studied 
	in detail. Based on the electronic structure evolution, we 
	calculated its switchable non-linear Hall performance under 
	a varying electric field. Taking advantage of the switchable 
	NHE and large current responsivity in monolayer $\text{TaIrTe}_4$, 
	a BCD based Hall rectifier could be designed and utilized in the 
	AC electrical fields and THz radiation detection.

	\section{Method}
	\subsection{Non-linear Hall effect}
	
	Consider two linear polarized THz radiations or AC electric fields 
	defined by $\bm{\varepsilon}_b(t)=\varepsilon_b e^{i\omega t}$ 
	and $\bm{\varepsilon}_c(t)=\varepsilon_c e^{i\omega t}$,
	where $b$ and $c$ are the polarized direction and $\omega$ is the 
	wave frequency.  The non-linear Hall rectify current along 
	$a$ direction is given by\cite{sodemann2015}
	\begin{equation}
		\begin{aligned}
			j_a = \chi_{a b c} \varepsilon_b \varepsilon_c^{*}
		\end{aligned}
	\end{equation}
	By considering the symmetry of this second-order tensor and ignoring 
	the conductivity coefficient, it's convenient to reform $\chi_{a b c}$ 
	into a two-indices quantity	$D_{b d}$. Besides, only $z$ component of 
	Berry curvature exists for 2D material. Thus, only components with 
	$d=z$ and $a,c=x,y$ can exist, which reads
	\begin{equation}
		\begin{aligned}
			\chi_{x b y} = -\chi_{y b x}  = \frac{e^3 \tau}{2 \hbar^2(1+i \omega \tau)} D_{b z}
		\end{aligned}
	\end{equation}
	
	\begin{equation}
		\begin{aligned}
			D_{b z}=\int [d \mathbf{k}] \sum_n f_{n\mathbf{k}} \frac{\partial \Omega_{z\mathbf{k}}^n}{\partial k_b} 
		\end{aligned}
	\end{equation}
	where $\tau$ is the electron relaxation time, and 
	$f_{n\mathbf{k}}=1/(e^{(E_{\mathbf{k}}-E_f-\mu)/\epsilon}+1)$ 
	is the Fermi-Dirac distribution under the equilibrium state.
	$\Omega_{z\mathbf{k}}^n$ is the Berry curvature of the $n$-th 
	band at momenta $\mathbf{k}$, which reads
	\begin{equation}
		\begin{aligned}
			\Omega_{z\mathbf{k}}^n=2 i \hbar^2 \sum_{m \neq n} \frac{\braket{u_{n\mathbf{k}}|\hat{v}_x|u_{m\mathbf{k}}}\braket{u_{m\mathbf{k}}|\hat{v}_y|u_{n\mathbf{k}}}}{\left(E_{n\mathbf{k}}-E_{m\mathbf{k}}\right)^2}
		\end{aligned}
	\end{equation}
	where $\hat{v}_a=\frac{1}{\hbar}\frac{\partial \hat{H}}{\partial k_a}$ 
	is the velocity operator along	$a$ direction, $\hat{H}$ is the 
	Hamiltonian in Wannier basis, $\ket{u_{n\mathbf{k}}}$ is the 
    eigenstate of Hamiltonian with the eigenvalue of $E_{n\mathbf{k}}$. 
	Note that the BCD is unit-free in three-dimension 
	and with a unit of Angstrom	in 2D.

	Considering the scattering rate of electrons is in 
	the order of tens of THz, the NHE in materials can be utilized
	in rectifying incident THz radiations with the output of direct
	electrical current. Corresponding performance of the THz detector 
	device can be characterized by the strength of the electric output 
	per optical input, which is known as the current repsonsivity\cite{zhang2021}.
	To obtain it, the power of the Hall rectifier $P$
	should also be taken into consideration other than the BCD,
	which reads
	\begin{equation}
		\begin{aligned}
			P = \frac{e \tau}{\hbar(1+i \omega \tau)} \sigma_{a b} \varepsilon_a \varepsilon_b LW
		\end{aligned}
	\end{equation}
	where $L$ and $W$ are the length and width of the Hall rectifier.
	$\sigma_{a b}$ is the Drude weight, which is given by 
	\begin{equation}
		\begin{aligned}
			\sigma_{a b}
			=&\frac{e}{\hbar}\int [d\mathbf{k}] \frac{\partial E}{\partial k_a} \frac{\partial E}{\partial k_b}\left(-\frac{\partial f_{\mathbf{k}}}{\partial E}\right)\\
			=&\frac{e}{\hbar}\int [d\mathbf{k}] \sum_n \braket{n|\hat{v}_a|n}\braket{n|\hat{v}_b|n}\delta(E_\mathbf{k}-E_n)
		\end{aligned}
		\label{eq:drude}
	\end{equation}
	Thus, the current responsivity $R_a$ of the Hall rectifier could 
	be given by the output current per adsorbed power which reads
	\begin{equation}
		\begin{aligned}
			R_a = \frac{I_H}{P} = \frac{1}{W}\frac{D_{az}}{\hbar\sigma_{aa}}
		\end{aligned}
	\end{equation}
	where the term of $\tau$/(1+$i\omega\tau$) with electron relaxation 
	time is canceled out.

	The monolayer $\text{TaIrTe}_4$ belongs to a centrosymmetric space 
	group $P2_1/m$, where nonlinear response is forbidden due to 
	the inversion symmetry.	After applying an electric field 
	perpendicular to the layer, this inversion symmetry is	broken
	and only a mirror symmetry \{$\mathcal{M}_{y}|(0,1/2,0)$\} is 
	preserved. This symmetry leads to an odd 
	$\frac{\partial \Omega_z}{\partial k_x}$ and an
	even $\frac{\partial \Omega_z}{\partial k_y}$. Therefore,
	only the components with $b=y$	
	($D_{yz}$, $\chi_{xyy}$, $\chi_{yyx}$, $\sigma_{yy}$ and $R_y$) 
	are allowed to exist. The numerical integration over the whole Brillouin 
	zone (BZ) is performed on a dense $k$-grid of 2000$\times$2000 and 
	500$\times$500 for the primitive cell and CDW phase of the monolayer  
	$\text{TaIrTe}_4$ respectively and are tested to be fully coverage for 
	both BCD and Drude term. Wilson loop is carried out in 
	Wannier basis to characterize the $Z_2$ topological charges
	of $\text{TaIrTe}_4$ under different electric fields\cite{yu2011}.

	\subsection{Ground state calculation}

	The crystal structure of the monolayer $\text{TaIrTe}_4$ is 
	fully optimized with force difference on each atom no more than 
	0.001 $\text{eV/\AA}$ in Vienna Ab Initio Simulation Package
	(VASP)\cite{kresse1996,kresse1996a}	and in good agreement 
	with experimental reports\cite{mar1992,tang2024}. Ground 
	state calculation is performed in Full Potential 
	Localized-orbital (FPLO) program to obtain a symmetry 
	conserved Hamiltonian in Wannier basis\cite{koepernik1999,opahle1999,koepernik2023,perdew1996}.
	The electric field is simulated by adding an on-site energy 
	which varies linearly with the Wannier center coordinates 
	perpendicular to the 2D plane on the Hamiltonian in Wannier 
	basis. The Hamiltonian of the 15$\times$1$\times$1 supercell 
	is built from the hopping strength extracted from the Wannier 
	function of the primitive cell. The CDW phase is simulated by 
	a superlattice modulation of Fröhlich-Peierls Hamiltonian 
	of $H=H_r+Vcos(Qx)u_r^{\dagger}u_r$, where $V$ is the 
	CDW modulation strength, $Q$ is the magnitude of the CDW vector 
	and $x$ is the Wannier center coordinate along $x$ 
	axis\cite{frohlich1954,yu2011}.

	\section{Results and Discussion}
	\subsection{Electronic structure of monolayer $\text{TaIrTe}_4$}
	
\begin{figure}[htbp]
		\begin{center}
			\includegraphics[width=0.48\textwidth]{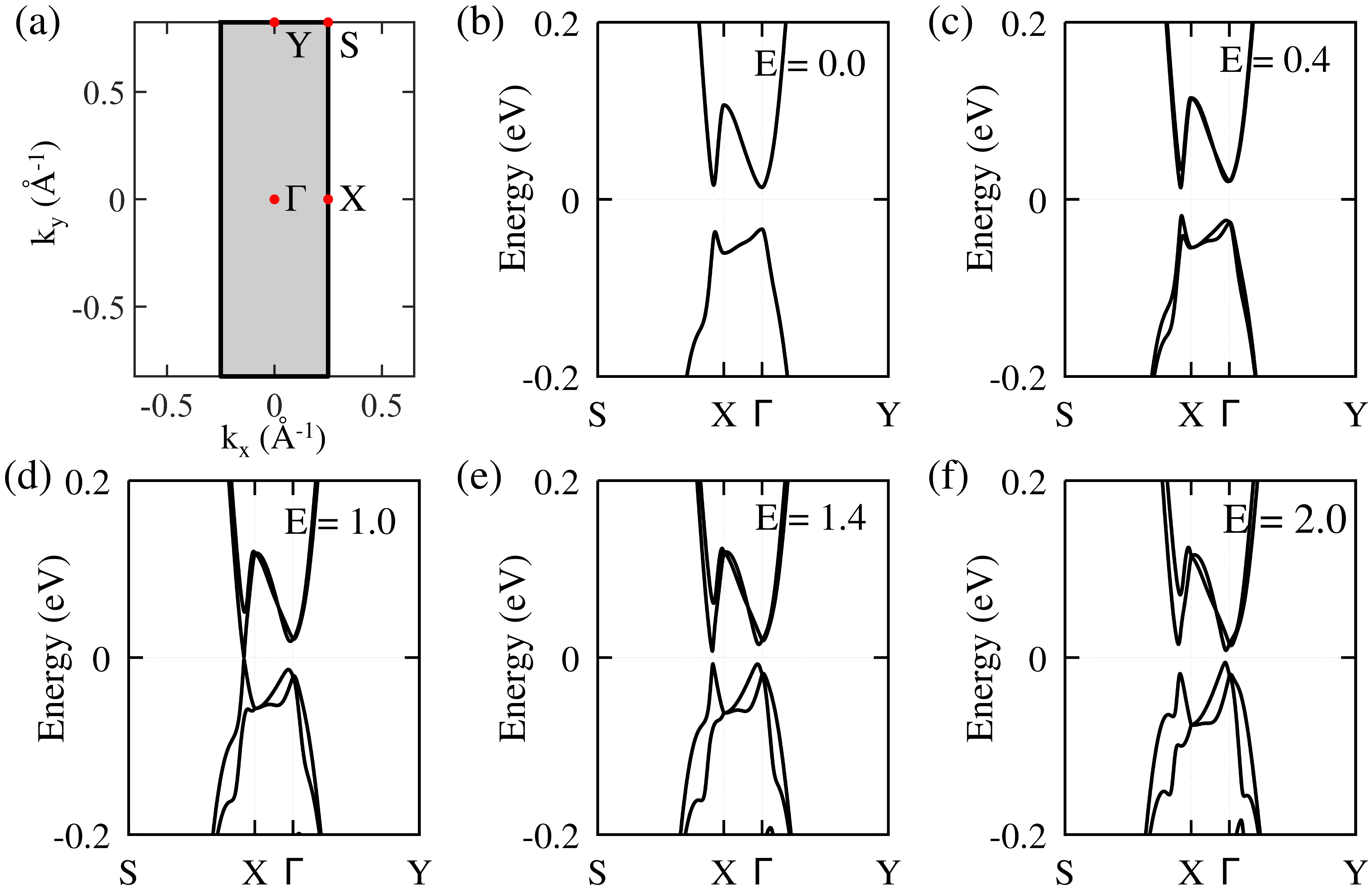}
		\end{center}
		\caption{
			(a) Brillouin Zone (BZ) of $\text{TaIrTe}_4$ monolayer;
			(b-f) Energy dispersions of $\text{TaIrTe}_4$ monolayer 
			with different electric fields $E$.
			$E$ is given in the unit of eV/c, with 
			$c=4.010~\text{\AA}$ as the 
			layer thickness of $\text{TaIrTe}_4$ monolayer. 
		}
		\label{fig2}
	\end{figure}

	The electronic band structure of monolayer $\text{TaIrTe}_4$ is 
	shown in Fig.~\ref{fig2}(b), where all the bands are doubly 
	degenerate due to the inversion symmetry. From energy dispersion 
	we can find a band inversion near $\Gamma$ point, which corresponds 
	to a QSHI phase with a band gap of around 45~meV, in agreement with 
	the previous reports\cite{liu2017a,guo2020,tang2024}.
	After applying an electric field along $z$, the inversion symmetry 
	is broken and leads to a band splitting between spin-up and 
	spin-down channels,	see Fig.~\ref{fig2}(c-f). This band splitting 
	gradually grows as the electric field increases. Meanwhile, the 
	band gap along $S$-$X$ gradually decreases and gets crossed when 
	the field reaches around 1.0~eV/c, see Fig.~\ref{fig2}(d). 
	After the crossing, the band gap is reformed and the monolayer 
	$\text{TaIrTe}_4$ becomes a trivial insulator with a band gap 
	of around 13~meV(E=2.0~eV/c).

	\subsection{Non-linear Hall effect in monolayer $\text{TaIrTe}_4$}
	
	\begin{figure}[htbp]
		\begin{center}
			\includegraphics[width=0.48\textwidth]{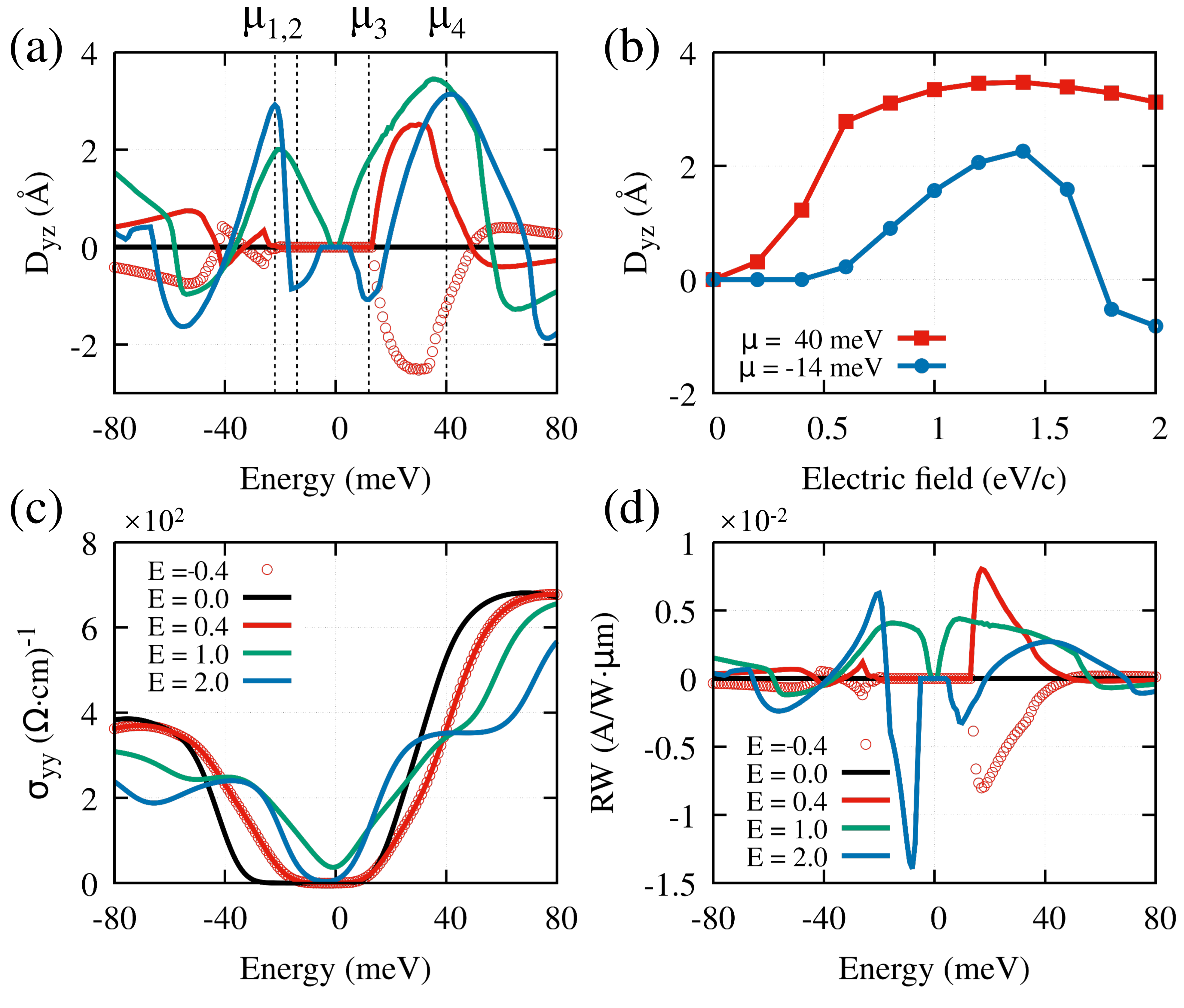}
		\end{center}
		\caption{(a) Chemical potential dependent BCD of 
		$\text{TaIrTe}_4$ monolayer at different electric fields.
			Four peaks can be found near the Fermi energy, with
			$\mu_1$=-22~meV, $\mu_2$=-14~meV, $\mu_3$=12~meV, $\mu_4$=40~meV.
			(b) The BCD of $\text{TaIrTe}_4$ monolayer at chemical potentials 
			of $\mu_2$ and $\mu_4$ with varying electric fields. 
			(c) The chemical potential dependent Drude weight 
			$\sigma_{yy}$ of $\text{TaIrTe}_4$ monolayer at different 
			electric fields.
			(d) The chemical potential dependent current responsivity 
			times width (RW) of $\text{TaIrTe}_4$ monolayer with
			different electric fields.
		}
		\label{fig3}
	\end{figure}

	The above band structure analysis demonstrates the inversion 
	symmetry broken after applying an out-of-plane field, 
	which is essential for generating a non-linear response.
	Besides, topological phase transitions would happen for 
	some given electric fields. Since such phase transitions 
	are often along with dramatic changes in Berry curvature 
	distribution, we analyzed corresponding non-linear Hall 
	transport in detail, see Fig.~\ref{fig3}. In the absence 
	of the electric field, the BCD is forced to zero due 
	to the inversion symmetry (black line). After applying a 
	small electric field of E = 0.4~eV/c, the $D_{yz}$ component 
	of BCD increases sharply from zero, consistent with our 
	symmetry analysis. Besides, a peak with the magnitude of  
	2.4~$\text{\AA}$ can be found at the chemical potential 
	of 40~meV (red line).

	As the field continually increases to 1.0~eV/c, a phase 
	transition from	QSHI to normal insulator happens. 
	Meanwhile, a new peak appears at around $\mu_1$
	and the peak at $\mu_4$ becomes larger (green line). 
	This can be attributed to the strength of inversion 
	symmetry breaking induced by the electric field. 
	After the phase transition when the electric field reaches
	around 2.0~eV/c (blue line), two peaks at $\mu_2$ and 
	$\mu_3$ appear with opposite signs with respect to that
	at $\mu_1$ and $\mu_4$.

	To have a further understanding of the tunability of BCD, 
	we analyzed the field dependent $D_{yz}$ at given chemical 
	potentials of $\mu_2$ and $\mu_4$, respectively, see 
	Fig.~\ref{fig3}(b). For the behavior away from Fermi 
	energy ($\mu_4$), the change of $D_{yz}$ is trivial as 
	it increases monotonically up to 1.0~eV/c and saturates 
	at around 3.5~$\text{\AA}$. Meanwhile, for the case near 
	Fermi energy ($\mu_2$), the rate of increase in $D_{yz}$ 
	slows down at 1.0 eV/c and subsequently decreases rapidly 
	after 1.4 eV/c, resulting in a sign change at 1.6 eV/c.
	This indicates the emergence of some negative contribution 
	after the topological phase transition. Besides, the 
	magnitude of this negative contribution increases much 
	faster than the positive counterpart with the increasing 
	electric field, resulting in a sign change near the Fermi energy.

	To obtain the absorbed power of the THz Hall rectifier, we also 
	calculated the Drude term $\sigma_{yy}$, see Fig.~\ref{fig3}(c). 
	Due to the intra-band Drude term, a zero plateau emerges 
	within the band gap near the Fermi level. The magnitude of 
	$\sigma_{yy}$ grows rapidly as it gets away from the gap.
	When an electric field is applied, the basic shape of $\sigma_{yy}$ 
	is generally kept,	and changes mainly occur in the zero plateau 
	range and the curve broadening. Thus, the current responsivity 
	R becomes larger near the band edge as it varies inversely with the 
	absorbed power $\sigma_{yy}$. This can be confirmed by the 
	analysis of the RW plot in Fig.~\ref{fig3}(d). Although the 
	magnitude of the BCD peak near $\mu_2$ is smaller compared 
	to the peak at $\mu_1$, the current responsivity exhibits 
	distinct behavior, as the peak near $\mu_1$ demonstrates 
	greater robustness and tunability.

	The intensity of RW could reach around 
	1.4$\times10^{-2}$ $\mu$m/V, indicating the potential to 
	get a current responsivity R of 0.14 V$^{-1}$ with a 
	sample width of 0.1 $\mu$m.	Taking advantage of the 
	controllable growth nature of 2D materials, a smaller 
	sample width can be achieved to yield a stronger signal. 
	Additionally, the direction of the nonlinear Hall current 
	can be flipped by the sign of the applied electric field, 
	as illustrated in the comparison between the red solid and 
	dot curves in Fig.~\ref{fig3}(a,b,d).

	\begin{figure}[htbp]
		\begin{center}
			\includegraphics[width=0.48\textwidth]{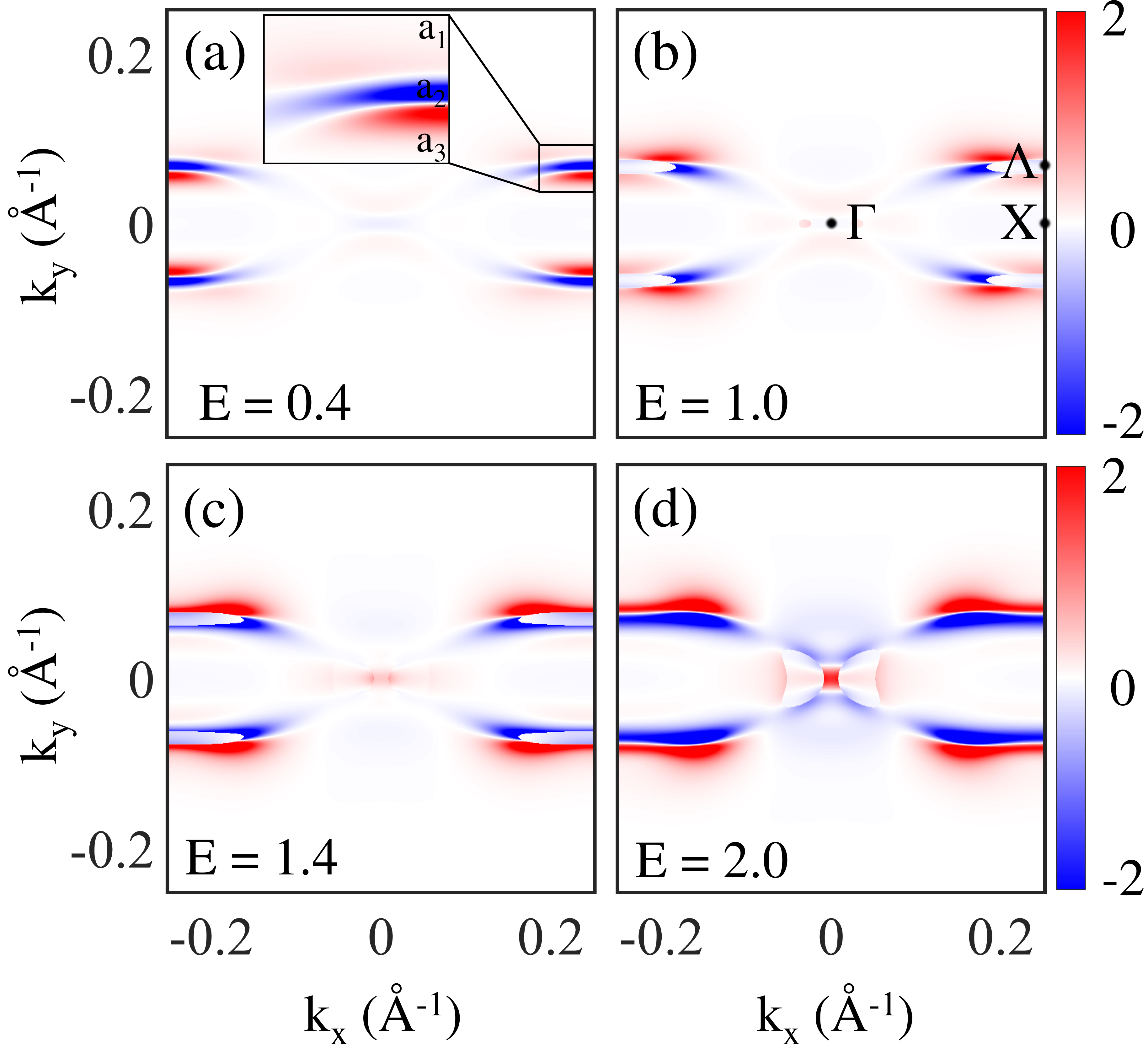}
		\end{center}
		\caption{
			The Berry curvature dipole distribution $\tilde{D}_{yz\bm{k}}=
			\sum_n \frac{f_{n\bm{k}}}{V}\frac{\partial \Omega_{z\bm{k}}^n}{\partial k_y}$ 
			under different electric fields of  
			(a) E = 0.4~eV/c, (b) E = 1.0~eV/c, (c) E = 1.4~eV/c, and (d) E = 2.0~eV/c
			with a given chemical potential $\mu_2$=-14~meV.
			BZ alone $k_y$ axis is zoomed in as the other part is negligible, 
			colorbars of $\tilde{D}_{yz\bm{k}}$ are all in unit of $10^4$.
		}
		\label{fig4}
		
	\end{figure}
	\begin{figure*}[htbp]
		\begin{center}
			\includegraphics[width=0.90\textwidth]{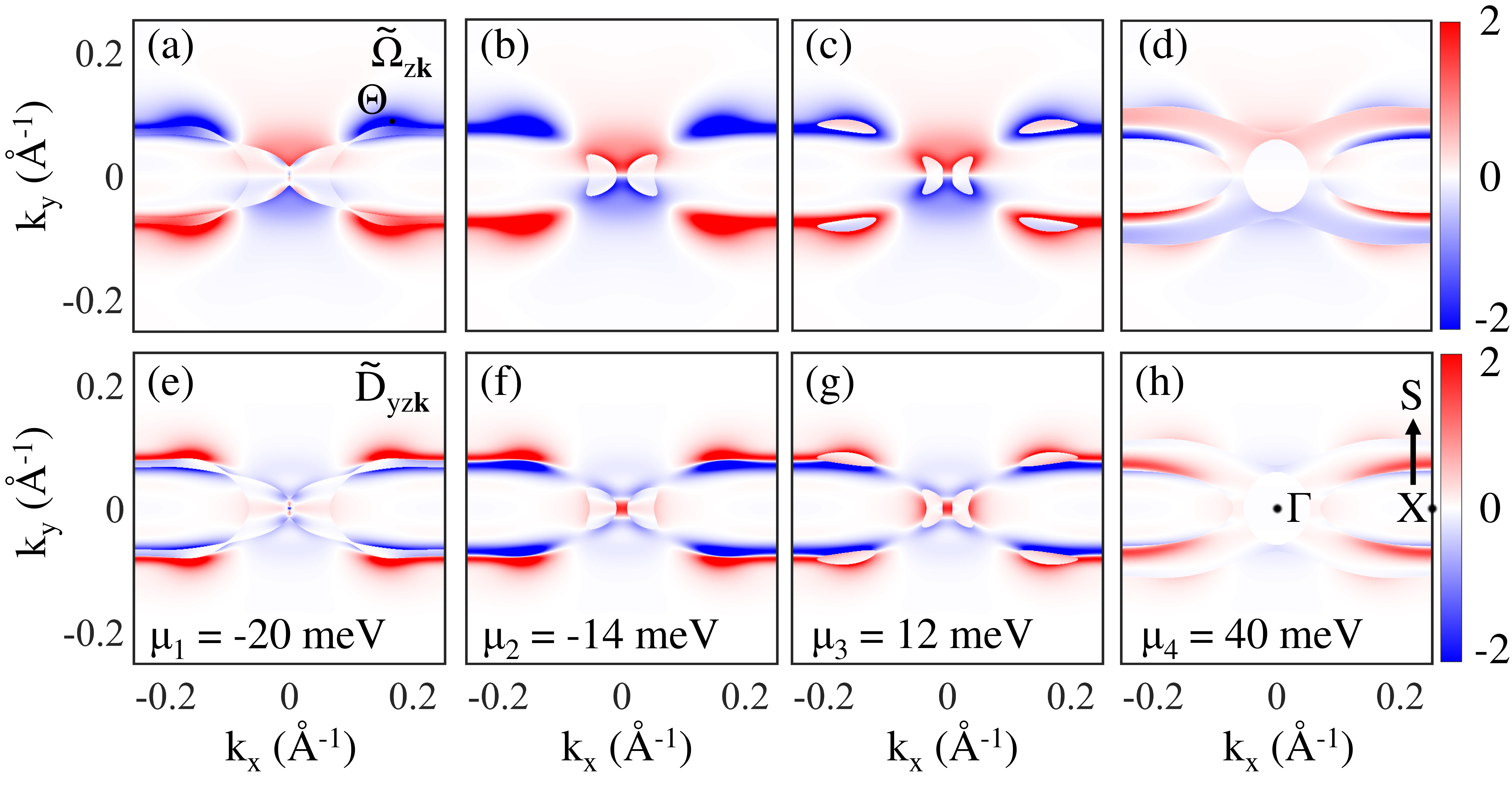}
		\end{center}
		\caption{
			The Berry curvature distribution $\tilde{\Omega}_{z\bm{k}}=\sum_{n}
			f_{n\bm{k}}\Omega_{z\bm{k}}^n$ of $\text{TaIrTe}_4$ monolayer
			at different chemical potentials of 
			(a) $\mu_1$=-22~meV, (b) $\mu_2$=-14~meV, (c) $\mu_3$=12~meV, 
			and (d) $\mu_4$=40~meV	under an electric field of E=2.0~eV/c. 
			The BCD distribution $\tilde{D}_{yz\bm{k}}$ of $\text{TaIrTe}_4$ 
			monolayer at different chemical potentials of 
			(e) $\mu_1$=-22~meV, (f) $\mu_2$=-14~meV, (g) $\mu_3$=12~meV, 
			and (h) $\mu_4$=40~meV	under an electric field of E=2.0~eV/c. 
			BZ along $k_y$ axis is zoomed in  as the 
			other part is negligible, color bars of 
			$\tilde{\Omega}_{z\bm{k}}$ and $\tilde{D}_{yz\bm{k}}$
			are in unit of $10^2~\hbar/e^2(\text{\AA})$ and $10^4$, 
			respectively.
		}
		\label{fig5}
	\end{figure*}
	
	\begin{figure}[htb]
		\centering
		\includegraphics[scale=0.11]{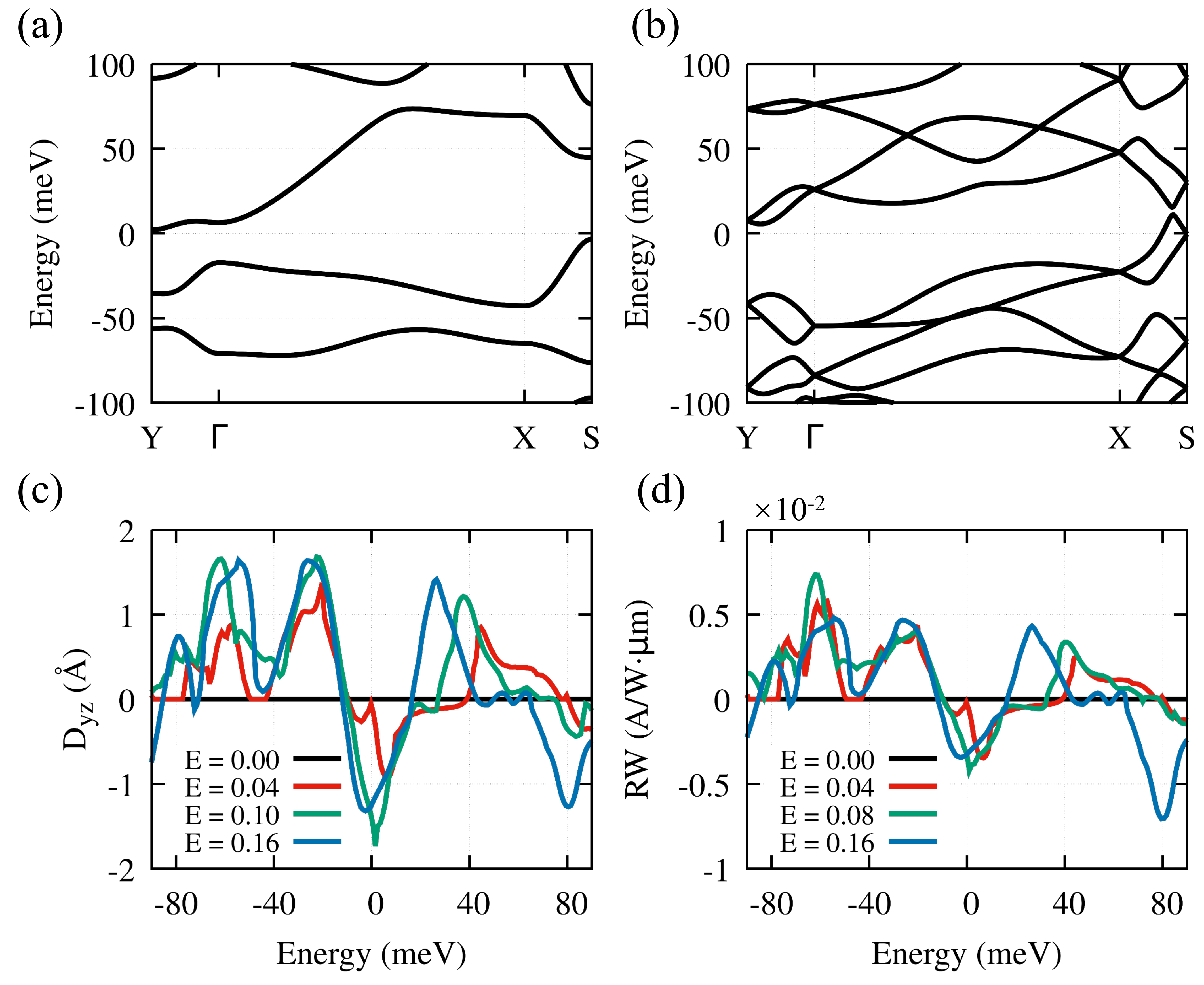}
		\caption{ 
			(a) Band structure of the CDW phase $\text{TaIrTe}_4$ monolayer.
			(b) Band structure of the CDW phase $\text{TaIrTe}_4$ monolayer
			under an electric field of E=0.16~eV/c.		
			(c) The chemical potential dependent BCD of 
			the CDW phase $\text{TaIrTe}_4$ monolayer
			at different electric fields.
			(d) The chemical potential dependent current responsivity times width (RW) of 
			the CDW phase $\text{TaIrTe}_4$ monolayer at different electric fields.
		}
		\label{fig6}
	\end{figure}%
	
	To understand the sign change of BCD under different
	electric fields and chemical potentials, we analyzed
	the local distribution $\tilde{D}_{yz\bm{k}}$ 
	and $\tilde{\Omega}^z_\mathbf{k}$, see Figs.~\ref{fig4} and \ref{fig5}.
	Since the Fermi crossing mainly takes place near the 
	$k_y=0$ line, the non-zero BCD distribution is confined 
	in the range of $k_y$ between -0.25 and 0.25 $\text{\AA}^{\text{-1}}$, 
	see Figs.~\ref{fig2} and \ref{fig4}. Applying an electric field, 
	a sign change at the chemical potential $\mu_2=-14~\text{meV}$ could 
	be found. The detailed local distribution of 
	$\tilde{D}_{yz\bm{k}}$ at $\mu_2=-14~\text{meV}$ is shown in 
	Fig.~\ref{fig4}. For the $\tilde{D}_{yz\bm{k}}$ under a 
	small electric field of E=0.4~eV/c, the high intensities
	mainly locate near the $\Lambda$ point, see Fig.~\ref{fig4}(a).
	
	For convenience, the dominant contribution is divided into 
	three areas around the $\Lambda$ point, where a$_1$ and a$_3$ 
	exhibit positive dominance while a$_2$ is negative.
	With detailed quantitative analysis, we can see that 
	a$_1$ and a$_3$ make a larger contribution compared to 
	a$_2$ at low electric field, see Fig.~\ref{fig4}(a),
	which results in the positive ${D}_{yz}$ in Fig.~\ref{fig3}(a).	
	As the electric field increases, the magnitude of $\tilde{D}_{yz\bm{k}}$ 
	in a$_1$ and a$_3$ grows and decays, respectively. 
	Meanwhile, the magnitude in a$_2$ increases rapidly and eventually 
	becomes dominated after the topological phase transition. 
	The joint effect of the three areas explains why the sign change 
	of $D_{yz}$ is later than the topological phase transition.

	The chemical potential-dependent BCD is also analyzed
	at a given E=2.0~eV/c, see Fig.~\ref{fig5}.
	Both local Berry curvature $\tilde{\Omega}_{z\bm{k}}$(Fig.~\ref{fig5}(a-d)) and 
	BCD $\tilde{D}_{yz\bm{k}}$ (Fig.~\ref{fig5}(e-h)) of the 
	four peaks in Fig.~\ref{fig3} are calculated.
	Owing to the mirror plane of \{$\mathcal{M}_{y}|(0,1/2,0)$\},
	$\tilde{\Omega}_{z\bm{k}}$ is even with respect to $k_x$ 
	and odd with respect to $k_y$  under an	out-of-plane electric field. 
	Since BCD is a first-order derivative of the Berry curvature, it exhibits 
	different parity characteristics compared to the Berry curvature with respect to momentum.
	Therefore, $\tilde{D}_{yz\bm{k}}$  presents as an even function of $k_x$ and $k_y$, respectively,
	see Fig.~\ref{fig5}(e-h).

	With the above understanding of symmetry analysis, we go to the details 
	about the Berry curvature distribution in $k$ space, see in Fig.~\ref{fig5}(a-d).
	As depicted in Fig.~\ref{fig5}(a), two hot spots could be seen
	around $\Theta$ and $\Gamma$ points in the positive BZ zone,
	which are negative and positive dominated, respectively. The three peaks of 
	$\mu_1$-$\mu_3$ in Fig.~\ref{fig3}(a) share similar Berry curvature 
	distributions, see the comparison of Fig.~\ref{fig5}(a-c). Owing to the 
	crossing of the conduction band, an empty part appears near the $\Theta$ point 
	in Fig.~\ref{fig5}(c). As a consequence, the distribution of 
	$\tilde{D}_{yz\bm{k}}$ in Fig.~\ref{fig5}(e-g)
	also shows a similar shape. The negative region in 
	$\tilde{\Omega}_{z\bm{k}}$ splits 
	into two areas with opposite signs in $\tilde{D}_{yz\bm{k}}$. 
	The intensity of these areas differs as the negative part is 
	stronger in $\mu_2$ and $\mu_3$, which primarily contributes 
	to the sign change of $D_{yz}$ near the Fermi level. Besides, the 
	magnitude of $\tilde{\Omega}_{z\bm{k}}$ around $\Gamma$  is
	small and makes little contribution to the $D_{yz}$.
	As presented in Figs.~\ref{fig5}(d) and (h), 
	there are multiple band contributions at $\mu_4$ = 40~meV, 
	and they are mainly dominated by the positive intensity, 
	leading to the corresponding peak in Fig.~\ref{fig3}(a).

	\subsection{Non-linear Hall transport in CDW phase}

	Apart from the QSHI state at the charge neutrality point, 
	a new $\text{Z}_2$ gap can be generated by placing the 
	Fermi level at van Hove singularity point. It has been 
	reported that the CDW phase in monolayer $\text{TaIrTe}_4$ has 
	a wave vector of Q*=(0.068,0)\cite{tang2024}. With the
	inspiration from experimental reports, we also computed the 
	BCD and corresponding current responsivity near the CDW gap. 
	According to the quantum electrical transport measurements, 
	a periodic potential of V=0.13~eV is taken into account 
	to generate the CDW phase, see Fig.~\ref{fig6}(a).	Similar to
	the original primitive cell, all bands 
	in Fig.~\ref{fig6}(a) are doubly degenerated due to the inversion 
	symmetry. Thus, an out-of-plane electric field is applied to 
	induce the non-linear Hall response, see Fig.~\ref{fig6}(b). 
	As a result, a significant band split was induced under a small 
	electric field of E=0.16~eV/c (approximately 0.4~eV/nm), see 
	Fig.~\ref{fig6}(b). All three CDW gaps disappear along with 
	the increasing of electric field.

	With the guiding principle from band structures, we analyzed the non-linear 
	Hall transport in detail. Multiple peaks of BCD and current responsivity 
	can be found in Fig.~\ref{fig6}(c,d) and they show similar trends under 
	different electric fields. There are three clear peaks near the Fermi level. 
	Two positive peaks with the magnitude of 1.67~\AA~ are located at around -54~meV and 
	-26~meV, respectively. While a negative peak of -1.31~\AA~ is located at around -3~meV. 
	These results suggest that the CDW phase of monolayer $\text{TaIrTe}_4$ 
	can also be tuned by shifting the chemical potential near the CDW gap.

	\section{Conclusion}
	In this work, we systemically studied the electronic structures and the 
	non-linear Hall transports in the new discovered topological state of 
	dual QSHI in $\text{TaIrTe}_4$.	
	Strong peaks of BCD were identified in the condition of a gating electric field
	and found to be switchable in terms of both sign and magnitude. 
	Through the analysis of momentum-resolved BCD, we reveal that the switchable 
	phenomenons correspond precisely to the topological phase transition
	in both the primitive cell and CDW phase.
    The results from our calculations illustrate that the topological
	band gaps in dual QSHIs offer natural platforms to obtain strong and tunable
	NHE and they exhibit promising potential applications in the next
	generation of THz radiation detection technologies.

	\begin{acknowledgments}
		This work is supported by the National Key R\&D Program of China 2021YFB3501503,
		National Natural Science Foundation of China (Grants No.52271016, No.52188101 ),
		and Liaoning Province (Grant No. XLYC2203080). Part of the numerical calculations 
		in this study were carried out on the ORISE Supercomputer(No. DFZX202319). 
	\end{acknowledgments}

	\appendix
	
	\bibliographystyle{ieeetr}

\end{document}